%% file: main.tex
\documentclass[sigconf]{acmart}
\usepackage{xspace}
\usepackage{longtable}
\usepackage{array}
\usepackage[table]{xcolor}

\usepackage{color}
\usepackage{nameref}
\usepackage{multirow}
\usepackage{tabularx} 
\usepackage{censor}
\usepackage{nameref}
\usepackage{enumitem}

\usepackage[most]{tcolorbox}

\usepackage{xcolor}          

\definecolor{lightblue}{RGB}{0, 0, 100}

\newtcolorbox{MyBox}{
  colback=white,
  colframe=lightblue,
  fonttitle=\bfseries,
  coltitle=black,
  sharp corners,
  boxrule=1pt,
  left=5pt,
  right=5pt,
  top=5pt,
  bottom=5pt,
  breakable
}

\usepackage{xcolor}
\usepackage{mdframed}

\newmdenv[
  backgroundcolor=black!6,
  leftline=true, rightline=false, topline=false, bottomline=false,
  linecolor=black, linewidth=3pt,
  innerleftmargin=10pt, innerrightmargin=10pt,
  innertopmargin=8pt, innerbottommargin=8pt,
  skipabove=8pt, skipbelow=8pt
]{rqbox}

\newcommand{\RQbox}[2]{%
\begin{rqbox}
\textbf{#1}\par\medskip
\textit{#2}
\end{rqbox}
}

\usepackage[most]{tcolorbox}
\usepackage{xcolor}

\newtcolorbox{databox}{
  colback=teal!20!white,   
  boxrule=0pt,             
  enhanced,
  borderline west={4pt}{0pt}{black}, 
  left=10pt,
  right=10pt,
  top=8pt,
  bottom=8pt,
  sharp corners
}

\usepackage{fontawesome5}

\newtcolorbox{resultbox}{
  enhanced,
  colback=blue!5,
  boxrule=0pt,
  sharp corners,
  borderline west={4pt}{0pt}{black},
  left=10pt,
  right=10pt,
  top=8pt,
  bottom=8pt
}

\AtBeginDocument{%
  }

\setcopyright{acmlicensed}
\copyrightyear{2016}
\acmYear{2016}
\acmDOI{XXXXXXX.XXXXXXX}
\acmBooktitle{ASE 2026}
\acmISBN{978-1-4503-XXXX-X/2018/06}

\begin{document}

\title{Software Project Management with LLM-Based Automation: Coordination, Validation, and Governance in Practice}


\author{Ronnie de Souza Santos}
\email{ronnie.desouzasantos@ucalgary.ca}
\affiliation{%
  \institution{University of Calgary}
  \city{Calgary}
  \state{Alberta}
  \country{Canada}
}

\author{Cleyton Magalhaes}
\email{cleyton.vanut@ufrpe.br}
\affiliation{%
  \institution{UFRPE}
  \city{Recife}
  \state{Pernambuco}
  \country{Brazil}}

\author{Italo Santos}
\email{isantos3@hawaii.edu}
\affiliation{%
  \institution{University of Hawai‘i at Mānoa}
  \city{Honolulu}
  \state{Hawaii}
  \country{USA}
  }

\begin{abstract}
\input{abstract}
\end{abstract}

\keywords{software project management, software managers, LLMs, case study}



\maketitle

\input{introduction}
\input{background}
\input{method}
\input{results}

\input{discussion}
 \vspace{-0.8cm}
\input{conclusion}

\nocite{*}

\bibliographystyle{ACM-Reference-Format}
\bibliography{bibliography}

\appendix

\end{document}

%% file: abstract.tex
As software engineering has evolved, development environments have increasingly integrated automated support for tasks such as testing, analysis, and code generation, requiring project management to coordinate both human work and automated workflows. In this paper, we investigate how software project managers perceive changes in their work and learning demands associated with LLM-based automation. Motivated by a gap in software engineering research, which has largely focused on task-level and developer-centered uses of LLMs, we adopted an exploratory case study approach to capture managerial perspectives on automation in practice. Based on the experience of software project managers working in a large, multi-project software organization, our analysis indicates that LLM-based automation influences planning, estimation, coordination, monitoring, and governance activities rather than introducing new formal management practices. Participants described LLMs as becoming embedded in everyday project work, producing uneven effects on productivity, increasing the need for review and validation, and reducing visibility into task execution. These effects contribute to greater reliance on managerial judgment and coordination. Learning demands were perceived as experiential and incremental, centered on understanding LLM capabilities and limitations, critically assessing generated artifacts, and guiding responsible use within teams. The findings provide empirical evidence on how software project management work is adapted in contexts where LLM-based automation is integrated into ongoing software development practice.

%% file: introduction.tex
\section{Introduction}
\label{sec:introduction}

Software project management is a key activity in software development, concerned with planning, coordinating, and guiding work toward the delivery of software systems within organizational and project constraints~\cite{jones2004software, procaccino2006software}. As software engineering has evolved, development environments have increasingly incorporated automated support for activities such as testing, analysis, and code generation, requiring managers to coordinate human work alongside tool-supported and partially automated workflows. Consequently, software project management remains a socio-technical practice that integrates planning, coordination, and organizational alignment while adapting to evolving development practices, including automation~\cite{oliveira2016software, kalliamvakou2017makes, gandomani2020role}. In this context, software project managers mediate organizational constraints, stakeholder expectations, and software development activities~\cite{procaccino2006software, shastri2021role}. Their responsibilities include integrating tools and development practices into coherent workflows while managing cost, schedule, scope, quality, and team coordination~\cite{oliveira2016software, gandomani2020role}. As automated development becomes more common, managers increasingly coordinate AI-assisted activities while maintaining accountability for project outcomes and software quality~\cite{procaccino2006software, shastri2021role}.

More recently, the literature reports an increasing presence of artificial intelligence (AI) and large language models (LLMs) in software development activities, including coding, documentation, testing, learning, estimation, monitoring, and coordination~\cite{hou2024large, rico2025challenges, assalaarachchi2025generative}. These technologies are increasingly used to automate or partially automate software engineering tasks, supporting information-intensive and generative activities while retaining human oversight and decision-making responsibility~\cite{cabrero2024can, assalaarachchi2025generative}. In practice, LLM-based automation is becoming embedded across multiple stages of the software lifecycle, influencing how work is performed, how artifacts are produced, and how software organizations ensure validation, accountability, and human oversight.

At the same time, existing work primarily emphasizes how software practitioners adapt their development routines to AI and LLM usage, with comparatively limited attention to how LLM-based automation is governed, validated, and integrated into real software projects~\cite{hou2024large, rico2025challenges}. While the literature reports growing adoption of AI-based tools for planning, reporting, and coordination~\cite{assalaarachchi2025generative}, limited empirical evidence explains how software project managers adapt their practices to establish governance, accountability, and human oversight when coordinating AI-assisted development. In particular, little is known about how managers respond to the variability, uncertainty, and coordination demands introduced by LLM-based automation in ongoing software projects. To address this gap, this study investigates how software project managers manage and interpret the integration of LLM-based automation in software development practice and poses the following research question:

\RQbox{RQ$_1$ -- Perceived Effects of LLMs in Software Project Management}
{How do software project managers perceive the changes and learning demands introduced by LLMs in software project management?}

To answer our RQ, we conducted an exploratory case study in a large software organization. We focused on the software project manager role as the unit of analysis and collected data through an online qualitative questionnaire administered to ten managers responsible for planning and coordinating ongoing projects within the organization. The instrument included open-ended questions designed to capture perceptions of managerial practices, challenges, and adaptation processes related to LLM usage. Responses were analyzed using thematic analysis complemented by descriptive statistics to support contextual interpretation. Our findings suggest that LLM-based automation is changing planning, estimation, coordination, monitoring, and governance rather than creating entirely new management practices. Participants described LLMs as increasingly integrated into everyday project work, introducing uneven productivity effects and requiring greater attention to review, validation, accountability, and managerial judgment. These findings indicate that trustworthy integration of LLM-based automation does not uniformly reduce effort but redistributes work across generation, validation, evaluation, and coordination activities. Learning demands were characterized as experiential and incremental, focusing on understanding LLM capabilities and limitations, critically evaluating generated artifacts, and supporting responsible use within teams. This paper makes the following contributions:

\begin{itemize}
    \item \textbf{An empirical characterization of governance and human oversight in LLM-based software development}, demonstrating that trustworthy adoption depends primarily on managerial judgment in planning, coordination, validation, and risk management rather than on new formal project management practices.
    
    \item \textbf{An identification of practices supporting trustworthy LLM-based automation}, showing how managers evaluate LLM-generated artifacts, guide responsible tool use, and address accountability and governance through everyday project management activities.
\end{itemize}

The remainder of this paper is organized as follows. Section~\ref{sec:background} presents the background of this study. Section~\ref{sec:method} describes the methodology. Section~\ref{sec:findings} reports the findings, which are discussed in Section~\ref{sec:discussion}. Finally, Section~\ref{sec:conclusions} summarizes the main contributions and concluding remarks.

%% file: background.tex
\section{Background}
\label{sec:background}

This section situates software project management and the use of LLMs within contemporary software engineering, providing context for understanding managerial challenges in AI-supported development environments. \\

\noindent\textbf{The Role of Software Managers.}
Early software engineering literature describes software managers as responsible for planning, estimation, progress tracking, quality control, and coordination in projects characterized by technical complexity and uncertainty~\cite{jones2004software, procaccino2006software}. As development practices evolved, software management became increasingly adaptive, relying on experience, negotiation, and informal coordination rather than strict adherence to predefined processes~\cite{kalliamvakou2017makes}. In agile and hybrid environments, managers continue to support prioritization, alignment, and coordination across teams and stakeholders~\cite{gandomani2020role, berenguer2024investigating}. More recent work characterizes software management as a socio-technical activity requiring communication, interpretation, accountability, organizational awareness, and decision making across organizational and technical boundaries~\cite{oliveira2016software, shastri2021role, peters2015educating}. \\

\noindent\textbf{LLMs in Software Engineering Practice.}
The literature reports the use of LLMs to support requirements engineering, implementation, debugging, documentation, testing, and learning activities~\cite{hou2024large, khojah2024beyond, santos2024we, zheng2025towards}. Rather than being associated with a single development phase, practitioners use LLMs opportunistically according to task demands and situational needs~\cite{meem2025software, champa2024chatgpt, krishna2024using}. Existing studies indicate that LLMs support exploration, clarification, and artifact generation while humans retain responsibility for interpretation, integration, and validation~\cite{coutinho2024role, santana2025software, rasnayaka2024empirical, santos2025model}. LLMs are also used for estimation, documentation synthesis, reporting, monitoring, and project coordination, extending their use beyond developer-facing activities~\cite{cabrero2024can, assalaarachchi2025generative, suwarno2025exploring, ali2026artificial}. However, prior work primarily characterizes LLMs as assistive resources, with comparatively less attention devoted to their role in software engineering automation~\cite{rico2025challenges, assalaarachchi2025generative}. \\

\noindent\textbf{LLMs for Software Engineering Automation.}
Beyond assisting practitioners, LLMs have been applied to automate code generation, program repair, testing, documentation, and requirements analysis, demonstrating their applicability across multiple software engineering activities~\cite{hou2024large, zhong2025large}. Existing work also reports workflow-level automation, where LLMs coordinate multiple development steps such as fault localization, patch generation, and validation within a single process~\cite{xia2025demystifying}. This extends automation from isolated tasks to coordinated workflows involving multiple artifacts and dependencies~\cite{hou2024large, xia2025demystifying}. At the same time, LLM-based automation introduces challenges related to evaluation, reliability, and control because generated artifacts require scalable assessment and human validation before integration into software projects~\cite{he2026llm, zhong2025large, rico2025challenges}. Although prior research has extensively investigated the technical capabilities of LLM-based automation, comparatively little empirical evidence explains how software managers coordinate, govern, and operationalize these automated workflows in ongoing software projects~\cite{rico2025challenges, hou2024large}.

%% file: method.tex
\section{Methodology} 
\label{sec:method}

We adopted an exploratory case study following established software engineering guidelines for empirical case studies~\cite{ralph2020empirical, runeson2009guidelines}. Case studies are appropriate for investigating contemporary phenomena in their real-world context, particularly socio-technical practices shaped by everyday work, interpretation, and coordination. Accordingly, this study investigates how software project managers perceive changes in their work and learning demands associated with the use of LLMs in software projects. From an automated software engineering perspective, the phenomenon under study is the integration of LLM-based automation into software development processes. Rather than evaluating a specific tool or technique, the study examines how LLM-based automation is incorporated, coordinated, and governed within existing development workflows. This reflects industrial practice, where automation is integrated into broader software engineering processes requiring planning, validation, and coordination across teams. Our study does not seek to compare tools or establish causal relationships. Instead, it aims to understand managerial adaptation and perceived challenges in a natural organizational setting, consistent with case study research that emphasizes contextual understanding and theoretical insight over statistical generalization~\cite{ralph2020empirical}. This perspective also supports identifying practical coordination and governance challenges associated with LLM-based automation in software projects.

\subsection{Case Study Context}

The case concerns a large, long-established software and innovation organization operating primarily in South America, with additional presence in North America and Europe. The organization functions as a research, development, and innovation center and delivers software-intensive projects for industrial clients across multiple sectors. The workforce includes software developers, quality assurance professionals, designers, data scientists, automation specialists, and professionals in software and technical management roles. The organization has approximately 1,200 employees, of which more than 70\% are directly involved in software development activities. These professionals are organized into more than 70 teams working across different development methodologies, including Scrum, Kanban, and Waterfall. At the time of the study, the organization was running at least 70 software projects concurrently. These projects covered several domains, including finance, healthcare, education, digital platforms, industrial systems, and data-driven services. Project work was organized around multidisciplinary teams, with management activities embedded in planning, coordination, monitoring, and decision-making across projects. The organization serves clients across multiple regions, including North America, Latin America, Europe, and Asia. 

The organization had already incorporated LLM-based tools into multiple stages of its development processes at the time of data collection, including activities such as code generation, documentation, testing support, and internal knowledge access. This context provides a setting in which LLM-based automation is not experimental, but part of ongoing project work, allowing observation of how automation interacts with established practices. We selected this organization as the case because it provides an analytically rich setting for studying software project management under conditions of increasing LLM adoption. The scale of operations, the diversity of domains, and the coexistence of many concurrent projects create conditions in which managerial practices, technical work, organizational constraints, and AI-mediated development are closely intertwined. These characteristics make the case suitable for an exploratory software engineering case study focused on socio-technical change and the practical integration of automated techniques.

\subsection{Participant Selection}

We defined the unit of analysis as the software project manager role. Within the organization, 57 individuals identified their primary role as software project manager or technical manager, being responsible for planning, coordinating, and supervising software projects and teams. From this bounded population, we applied random sampling and contacted 25 managers to invite them to participate in the study. Participation was voluntary, and 10 managers completed the questionnaire. Although the number of respondents is limited, these participants were collectively responsible for at least 30 of the 70 projects active in the organization at the time of data collection, representing a substantial portion of ongoing project activity within the case. Given that software project managers typically supervise multiple projects simultaneously, the participating managers provide visibility into a significant share of the organization’s development work. This perspective allows the study to capture managerial practices across multiple teams, domains, and project contexts within the same organizational setting. From an automated software engineering perspective, the relevance of this sample lies in its coverage of projects in which LLM-based automation is actively used. Rather than sampling a large number of individuals with limited exposure to automation, the study focuses on a smaller group of participants who collectively oversee a wide range of projects where automation is embedded in practice. This increases the analytical relevance of the data for understanding how automated techniques are managed in real-world settings.

This sampling strategy sought to balance feasibility with analytical relevance. Rather than aiming for statistical representativeness, the study prioritized access to practitioners directly involved in managing projects influenced by LLM adoption. Consequently, the resulting sample constitutes a self-selected qualitative sample suitable for exploratory case study research in software engineering, where contextual insight and information richness are central concerns. In qualitative research, sample size is determined by the depth and richness of the data rather than by statistical requirements. Prior work indicates that meaningful thematic patterns can emerge with relatively small samples, particularly in specialized populations, and that data saturation is often observed within a range of approximately six to twelve participants~\cite{guest2006many}. Given the specialized nature of the software project manager role and the bounded organizational context, the sample size is consistent with established qualitative research practices.

\subsection{Pilot}

We validated the questionnaire through a pilot study with two researchers experienced in software management research. Feedback from the pilot led to improvements in wording clarity, refinement of concepts related to LLM capabilities and limitations reported in prior work, and a reduction in overlap among questions addressing planning, learning, and governance. Table~\ref{tab:survey} presents the full instrument. Data collection happened during January 2026.

\subsection{Data Collection}

We initially planned to collect data through semi-structured interviews, which are commonly used in software engineering research to capture practitioners’ experiences and interpretations of managerial work and decision-making. In qualitative research, interviews are often preferred because they allow probing, clarification, and deeper exploration of participants’ responses~\cite{seaman1999qualitative, guest2006many, ralph2020empirical}. From the perspective of studying automation in practice, interviews are particularly useful for understanding how practitioners interpret and interact with automated systems, including how they validate outputs, manage uncertainty, and integrate automated artifacts into workflows. However, during recruitment, we observed that this approach was impractical given the professional profile of the targeted participants. Managers reported highly restricted availability, which limited the feasibility of synchronous interviews. We therefore transformed the interview guide into an online qualitative questionnaire, following prior software engineering research that adapts qualitative instruments to accommodate participants’ time constraints while preserving open-ended reflection. This adaptation allowed participants to respond asynchronously while preserving the reflective and open-ended nature of the original questions. 

Although this approach reduces opportunities for real-time probing, it enabled access to participants who would otherwise be difficult to reach and is considered an acceptable alternative in qualitative studies involving time-constrained professional populations~\cite{seaman1999qualitative}. The questionnaire was designed to elicit detailed accounts of managerial practices, perceived changes, and learning demands associated with LLM usage in software projects. In particular, questions were structured to capture how managers plan, coordinate, monitor, and validate work in contexts where automated outputs are generated by LLMs. The questions were developed based on prior work on LLM usage in software engineering and on emerging discussions of AI-supported activities in software project management~\cite{hou2024large, assalaarachchi2025generative, rico2025challenges}. 

\input{tables/quest}

\subsection{Data Analysis}

The questionnaire yielded both qualitative and quantitative data. We analyzed open-ended responses using thematic analysis~\cite{terry2017thematic}, while we summarized demographic and contextual data using descriptive statistics~\cite{george2018descriptive}. We conducted thematic analysis following procedures commonly recommended for qualitative research in software engineering~\cite{seaman1999qualitative}.  From the perspective of automated software engineering, the analysis focused on identifying how LLM-based automation influences managerial activities such as planning, estimation, coordination, monitoring, and governance. The goal was not to evaluate the performance of the automation itself, but to characterize how its presence affects decision-making, validation effort, and coordination practices in real projects.

We coded the received answers following iterative cycles. In the first cycle, we read each response carefully and assigned codes capturing the core idea expressed by the participant. In the second cycle, we grouped conceptually related codes into broader categories representing recurring perceptions, challenges, and learning demands. We documented a rationale for each category to clarify its meaning and analytical scope. A final refinement cycle focused on ensuring conceptual clarity and consistency across themes. One researcher conducted the initial coding. A second researcher reviewed the codes and categories. We discussed the disagreements until we reached a shared interpretation. This process was used to enhance credibility, transparency, and reflexivity, and to reduce the influence of individual researcher assumptions.

Although the primary emphasis of the analysis was qualitative, we used descriptive statistics to summarize participant characteristics and to indicate the frequency of selected response options and themes. These summaries support contextual interpretation rather than inferential claims. In the context of this study, descriptive statistics are used to complement qualitative insights by indicating how frequently certain perceptions or challenges appear across participants, rather than to support generalization beyond the case.

\subsection{Ethics}
We conducted the study in accordance with institutional guidelines for research involving human participants. We obtained informed consent electronically at the beginning of the questionnaire. We informed participants about the study objectives, voluntary participation, anonymity, and their right to withdraw before submission. We did not collect personally identifiable information. We analyzed all data in aggregated form and used them exclusively for academic research purposes.

\subsection{Threats to Validity} 
As with any qualitative case study, this study is subject to limitations inherent to the method~\cite{runeson2009guidelines, ralph2020empirical}. First, the findings are based on ten software project managers. While this number may appear limited from a quantitative perspective, qualitative research prioritizes depth and contextual richness over sample size~\cite{ralph2020empirical}, and prior work reports credible findings with comparable or smaller samples~\cite{baker2012many, elmholdt2026many}. Participants were drawn from a bounded population of fifty-seven managers and collectively managed approximately thirty projects, representing about 42\% of ongoing projects in the organization. Although not statistically representative, this provides analytical representation of managerial practice within the case. Second, this study does not aim for statistical generalization but follows an analytic, transfer-oriented approach~\cite{ralph2020empirical}. We provide detailed context, participant characteristics, and methodological transparency, along with a replication package to support reanalysis and comparison. While we do not claim direct generalization, the level of detail enables assessment of applicability in other settings. Third, the analysis relies on self-reported accounts collected through an online qualitative questionnaire, reflecting perceptions rather than direct observation. This is common in studies of managerial work. To mitigate this limitation, we grounded findings in recurring patterns and supported interpretations with quotations. Finally, the study does not include quantitative measurement or causal inference, consistent with its exploratory design. It should be evaluated using qualitative criteria such as credibility, transparency, and transferability~\cite{ralph2020empirical}, within which it provides a systematic and empirically grounded account of LLM-based automation in practice.

%% file: tables/quest.tex
\begin{table}[t]
\caption{Survey Questionnaire}
\label{tab:survey}
\centering
\small
\begin{tabular}{p{7.5cm}}
\hline

\textbf{LLMs and Software Management} \\

1. Which skills are most important for planning and estimation in software projects where LLMs are used? \\

2. How have you developed or adapted your management skills due to LLM adoption? \\

3. Which control and monitoring skills have become more important with LLM usage? \\

4. Which challenges best reflect your experience managing projects involving LLMs (e.g., planning, monitoring, coordination, decision making, and risk management)? \\

\hline

\textbf{Learning and LLMs} \\

5. Which new skills have you needed to learn to manage LLM-mediated software projects? \\

6. Which area has presented the greatest managerial challenge (e.g., technical understanding, quality assurance, leadership, governance, or continuous learning)? \\

\hline

\textbf{Governance} \\

7. Which governance, quality, or responsibility decisions related to LLM usage have you made? \\

8. Which governance aspect has required the greatest managerial effort (e.g., quality assurance, acceptable use, responsibility, or compliance)? \\

9. How are decisions related to LLM usage currently handled in your projects? \\

\hline

\textbf{Participant Profile} \\

10. Management experience in software projects. \\

11. Number of software projects currently supervised. \\

12. Approximate number of people under responsibility. \\

13. Frequency of LLM usage within the team. \\

14. Main software engineering activities supported by LLMs. \\

15. Gender identity. \\

\hline
\end{tabular}
\end{table}

%% file: results.tex
\section{Results} 
\label{sec:findings}

Our sample comprised ten software project managers with diverse managerial backgrounds and responsibilities (Table~\ref{tab:participants}). Overall, participants represented a range of management experience, project workloads, and team sizes, providing perspectives from managers overseeing different organizational contexts. LLM usage was well established across participants' projects (shown below). The reported uses concentrated primarily on development and quality assurance activities, particularly test generation, code writing or review, technical problem resolution, and documentation support, with less frequent use for design exploration, requirements-related work, and experimental code generation.
\input{tables/participants}

\begin{resultbox}
\label{res:llmusage}
\faChartBar\ \textbf{Overview: LLM usage across participants’ projects}
\small
\medskip
\textbf{Frequency of LLM usage in projects}

\begin{center}
\begin{tabular}{lc}
\hline
Usage frequency & Participants \\
\hline
Always used & 5 \\
Frequently used & 4 \\
Occasionally used & 1 \\
\hline
\end{tabular}
\end{center}

\medskip
\textbf{Main activities supported by LLMs}

\begin{center}
\begin{tabular}{lc}
\hline
Activity & Mentions \\
\hline
Test generation & 8 \\
Code writing/review & 7 \\
Technical problem resolution & 6 \\
Documentation support & 5 \\
Design exploration & 3 \\
Requirements clarification & 3 \\
Experimental code generation & 1 \\
\hline
\end{tabular}
\end{center}

\medskip
LLM usage is embedded mainly in development and quality-related tasks, with more limited adoption in design and requirements activities.
\end{resultbox}

\subsection{Perceived Changes in Software Project Management Practices with LLM Usage}

Participants described planning and estimation in projects using LLMs as becoming less predictable and more dependent on managerial judgment. Rather than assuming uniform productivity gains, managers reported interpreting the effects of LLMs according to the task and development context. As one participant noted, \textit{“while some tasks may be accelerated, others require more time for review and validation”} (P003), while the same participant emphasized the need to avoid \textit{“unrealistic productivity expectations”} (P003). These accounts indicate that LLM-based automation introduces variability that must be interpreted in context. Planning and estimation also increasingly incorporate effort related to reviewing, validating, and integrating LLM-generated artifacts. One participant referred to \textit{“...explicitly considering the additional effort required for review, validation, and integration”} (P003), while another noted that \textit{“Some LLMs result in faster development, but many contribute primarily to improved quality, without necessarily accelerating the development process”} (P002). These observations suggest that LLM-based automation redistributes effort toward evaluation and integration rather than uniformly reducing development effort.

Participants also described coordination and communication as becoming more important for aligning individual LLM use with collective development workflows. One participant emphasized \textit{“Strong communication is also important to introduce and share initiatives related to LLM use within the broader project context with the team”} (P001), while another referred to \textit{“Coordination between human work and the use of AI tools, due to the need to align human contributions and LLM generated artifacts within the collective project workflow”} (P003). Managers further reported reduced visibility into work mediated by LLM interactions, making monitoring and control more difficult. One participant referred to \textit{“Progress and productivity monitoring, due to reduced visibility into how work is carried out when part of the tasks is mediated by individual interactions with LLMs”} (P001), while another highlighted \textit{“progress and productivity monitoring, due to reduced visibility into how work is performed when tasks are mediated by individual interactions with LLMs”} (P002). Despite these changes, participants emphasized continuity with established practices. One participant noted that \textit{“Since the adoption of these tools is recent, I still use standard estimation practices”} (P005), while another emphasized that \textit{“The same skills established before the LLM era remain relevant”} (P008). These accounts suggest that LLM-based automation extends existing management practices while increasing reliance on managerial judgment.

Finally, participants described greater uncertainty in project execution due to variability in LLM outputs and their impact on deadlines, quality, and rework. One participant described the need for \textit{“Risk management and decision making under uncertainty, since decisions must be made even when the impact of LLM usage on deadlines, quality, or rework is not predictable”} (P001), while another referred to \textit{“decision making with incomplete or variable information, due to non deterministic outputs, frequent shifts in approach, and limited LLM explainability”} (P002). These accounts indicate that managers increasingly rely on continuous interpretation and adaptation when coordinating LLM-supported projects.

\subsection{Learning Demands and Skill Adaptation in LLM-Mediated Software Project Management}

Participants described learning as emerging primarily through direct interaction with LLM-based automation rather than formal training. Managers emphasized developing an understanding of LLM capabilities and limitations through practical experience. One participant highlighted the need to \textit{“Understanding what LLMs are and knowing where and how the team began to apply them in projects”} (P001), while another stated that \textit{“The first step was to use the tools myself”} (P006). These accounts indicate that learning is closely tied to the practical integration of automation into everyday work. They also emphasized the need to critically evaluate LLM-generated outputs and adapt planning practices based on observed outcomes. One participant described the need to \textit{“evaluate critically the results generated by the tools, understanding when they really help and when they can introduce errors or inconsistencies”} (P003), while another noted that teams may \textit{“struggle to maintain adequate review and quality disciplines due to the convenience and volume of artifacts generated”} (P008). Planning and estimation were likewise adapted to account for validation and integration effort, as illustrated by \textit{“...explicitly considering the additional effort required for review, validation, and integration”} (P003). These accounts suggest that managers learn primarily by observing the practical effects of LLM-based automation.

Learning also involved guiding teams in the responsible use of LLMs by aligning expectations and encouraging appropriate use. One participant emphasized the need to \textit{“align expectations around responsible LLM use”} (P003), while another described their role as \textit{“reinforce the importance of their use and to demonstrate the positive outcomes of adopting them within the team”} (P005). Participants also referred to technical and organizational aspects of LLM adoption, including \textit{“Prompt writing and specification files for code generation, configuration of tools for automated review”} (P008), \textit{“Legal considerations regarding how far AI could be used”} (P004), and situations where \textit{“due to confidentiality issues in the projects, the use of AI is limited”} (P009). These accounts indicate that learning encompasses both enabling and constraining the use of LLM-based automation. Participants discussed that these learning demands extend rather than replace existing managerial competencies. One participant stated that \textit{“the fundamental skills remain the same, if not even more required”} (P008), while another noted that \textit{“For now, I continue to rely on traditional management skills”} (P010). These observations suggest that LLM-based automation leads to incremental skill adaptation rather than fundamental role transformation.

\subsection{Governance, Responsibility, and Managerial Decision Making in LLM-Mediated Projects}

Governance, responsibility, and accountability emerged as recurring concerns in projects using LLM-based automation. Managers described uncertainty regarding data usage, artifact authorship, compliance, and responsibility in case of failure. A central governance challenge concerned ensuring the quality and reliability of generated artifacts. One participant highlighted the need for \textit{“Ensuring the quality and reliability of generated artifacts, since the origin and degree of dependence on LLMs are not always transparent”} (P003). This illustrates the need to validate generated outputs before they are incorporated into project workflows.

Managers also described decisions related to defining acceptable LLM usage and aligning expectations across teams. One participant explained the need to \textit{“It was necessary to define which activities allowed free use of LLMs and which required more rigorous review”} (P003), while another referred to \textit{“Decisions about which data could be provided to AI tools and whether individuals were appropriately reviewing and correcting code generated by AI”} (P004). These accounts show that LLM adoption introduces governance decisions regarding acceptable use, review practices, and organizational boundaries. Responsibility attribution also emerged as an important concern. One participant emphasized that \textit{“any result generated by LLMs remained the responsibility of the team rather than the tool”} (P003), reinforcing that accountability continues to reside with human practitioners.

Governance was generally described as relying on managerial judgment rather than formal organizational frameworks. Participants reported that \textit{“They are handled on a case by case basis, as specific situations arise”} (P002). In contrast, some managers indicated that governance demands had remained limited, stating \textit{“So far, I have not needed to make specific decisions related to the use of LLMs in the projects I manage”} (P010). Together, these accounts suggest that governance currently depends primarily on contextual managerial decisions rather than standardized organizational processes.

\subsection{Answering RQ}
Software project managers perceived LLM-based automation as introducing selective changes to project management by increasing variability, validation effort, coordination demands, and reliance on managerial judgment, while learning remained incremental, governance was handled primarily through case-by-case decisions, and established management practices continued to underpin project coordination and control.


\begin{resultbox}
\faChartBar\ \textbf{RQ -- Perceived Effects of LLMs in Software Project Management}
\small
\medskip
\textbf{Perceived changes in project management}

\begin{center}
\begin{tabular}{p{0.9\linewidth}}
\hline
Reported perception \\
\hline
Selective and uneven changes rather than a full transformation of managerial roles \\

Greater reliance on managerial judgment and contextual interpretation of productivity gains \\

Increased attention to coordination, visibility, and risk monitoring \\

Many established management practices and skills remain central \\
\hline
\end{tabular}
\end{center}

\medskip
\textbf{Emerging learning demands}

\begin{center}
\begin{tabular}{p{0.9\linewidth}}
\hline
Managerial response \\
\hline
Incremental and experience-based learning about LLM capabilities and limitations \\

Need to critically evaluate LLM-generated artifacts \\

Managers support responsible adoption rather than replace existing competencies \\
\hline
\end{tabular}
\end{center}

\medskip
\textbf{Governance implications}

\begin{center}
\begin{tabular}{p{0.9\linewidth}}
\hline
Observed pattern \\
\hline
Responsibility management handled through managerial judgment and expectation alignment \\

Governance typically informal and addressed case by case rather than through structured frameworks \\
\hline
\end{tabular}
\end{center}

\medskip
\end{resultbox}

%% file: tables/participants.tex
\begin{table}[!ht]
\caption{Participant Demographics and Responsibilities}
\label{tab:participants}
\centering
\small
\begin{tabular}{l|l}

\hline
\textbf{Characteristic} & \textbf{Distribution} \\
\hline \hline
Gender identity &
4 women; 5 men; 1 undisclosed \\ \hline

Managerial experience &
4 (2--5 years); 3 (6--10 years); \\
& 3 ($>$10 years) \\ \hline

Projects supervised &
2 (1 project) 6 (2--3 projects);  \\
& 2 (4--5 projects); \\ \hline

People supervised &
3 (6--10 people) 2 (11--20 people); \\
& 5 ($>$20 people);  \\ \hline
\hline
\end{tabular}
\end{table}

%% file: discussion.tex
\section{Discussion} \label{sec:discussion}

\noindent \textbf{Comparison with the Literature.}
Prior software engineering research has primarily investigated the use of LLMs for code generation, debugging, documentation, testing, learning, and other task-level activities~\cite{hou2024large, khojah2024beyond, santos2024we, zheng2025towards}. More recent studies also describe collaborative uses, characterizing LLMs as supporting exploration, clarification, and artifact generation while humans retain responsibility for interpretation, validation, and integration~\cite{coutinho2024role, rasnayaka2024empirical, santos2025model}. Our findings are consistent with this literature regarding usage patterns but extend it by showing how these task-level capabilities translate into project-level effects. Rather than remaining individual activities, validation, interpretation, and tool usage become coordination challenges that affect planning, estimation, monitoring, and governance.

Participants described LLM-based automation as introducing variability in task execution, uneven productivity effects, and reduced visibility into development work, increasing reliance on managerial judgment and contextual interpretation. This is consistent with socio-technical perspectives of software project management that emphasize judgment, negotiation, and decision making under partial information~\cite{procaccino2006software, shastri2021role}. Rather than introducing fundamentally new management practices, LLM-based automation appears to intensify these existing demands.

The findings also position learning and governance as central managerial concerns. While prior work often emphasizes individual skills such as prompt writing and tool proficiency, participants described learning as developing capabilities for evaluating generated artifacts, aligning expectations across teams, and defining acceptable use boundaries. Governance likewise relied on ongoing managerial judgment regarding quality assurance, responsibility attribution, and compliance. Together, these findings suggest that LLM-based automation affects not only task execution but also how responsibility, coordination, and accountability are organized within software projects.
\\

\noindent \textbf{Implications.}
From a research perspective, this study contributes to automated software engineering by extending the focus from task-level outcomes to project-level coordination and control. The findings indicate that LLM-based automation introduces variability, redistributes effort toward validation and integration, and reduces visibility into work processes, suggesting that task-level performance alone provides only a partial view of its impact. Future research should investigate how these effects vary across organizational contexts, project criticality, and regulatory environments, as well as how coordination, learning, and governance evolve as LLM adoption matures. From a practice perspective, the findings indicate that LLM-based automation requires adjustments to project planning, monitoring, and governance. Based on the results, we derive the following recommendations:

\begin{itemize}
    \item \textbf{Adjust estimation practices to account for variability in automated outputs.} Because LLM effects vary across tasks and roles, estimation practices should account for variability rather than assume uniform productivity gains.

    \item \textbf{Incorporate explicit review responsibilities into team roles.} Review and validation responsibilities should be explicitly assigned to maintain quality and consistency, particularly among less experienced practitioners.

    \item \textbf{Address uneven impacts of automation across experience levels.} Project management practices should consider differences in practitioner experience when assigning tasks and interpreting LLM-generated outputs.

    \item \textbf{Align automation use with project constraints such as confidentiality and compliance.} Decisions regarding LLM adoption should reflect project-specific constraints, particularly when handling sensitive or regulated information.

    \item \textbf{Monitor the balance between speed and quality in LLM-mediated work.} Managers should monitor both delivery speed and output quality, recognizing that productivity gains may require additional validation effort.
\end{itemize}

These implications suggest that LLM-based automation should be integrated through appropriate coordination, validation, and governance mechanisms rather than being viewed solely as a means of increasing individual productivity. \\

%% file: conclusion.tex
\section{Conclusions and Future Work} 
\label{sec:conclusions}

Our study focused on how software project managers perceive changes in their work and learning demands associated with the use of LLMs in software projects. Using an exploratory case study, we investigated managerial sense-making, adaptation, and responsibility in contexts in which LLMs are embedded in everyday development and coordination activities. In general, our study aimed to understand how managers interpret and respond to the use of LLMs within ongoing software project management practice. Our findings indicate that managers perceive LLMs as becoming embedded in existing project management work rather than as introducing new roles or formal management approaches. LLM adoption does not reduce the importance of management, nor does it automate managerial work. Instead, it changes the conditions under which management is performed, increasing attention to uncertainty, coordination, learning, and accountability while remaining embedded in established project management practices. For \textbf{future work}, we intend to expand the body of knowledge generated in this study by administering a survey to software project managers to assess the prevalence and variation of the identified patterns across a broader range of organizational contexts. In addition, we plan to develop training materials for software managers based on the practices, learning needs, and governance challenges identified in this study, thereby supporting more informed and responsible use of LLMs in software project management.


\begin{databox}
\textbf{Data availability statement:}
To ensure verifiability and replicability, we provide the replication package of this study available at~\url{https://figshare.com/s/9e721bea3d114f7b9a24}.
\end{databox}



%% file: bibliography.bib
@article{kalliamvakou2017makes,
  title={What makes a great manager of software engineers?},
  author={Kalliamvakou, Eirini and Bird, Christian and Zimmermann, Thomas and Begel, Andrew and DeLine, Robert and German, Daniel M},
  journal={IEEE Transactions on Software Engineering},
  volume={45},
  number={1},
  pages={87--106},
  year={2017},
  publisher={IEEE}
}

@article{gandomani2020role,
  title={The role of project manager in agile software teams: A systematic literature review},
  author={Gandomani, Taghi Javdani and Tavakoli, Zeinab and Zulzalil, Hazura and Farsani, Hadi Khosravi},
  journal={IEEE access},
  volume={8},
  pages={117109--117121},
  year={2020},
  publisher={IEEE}
}

@article{procaccino2006software,
  title={Software project managers and project success: An exploratory study},
  author={Procaccino, J Drew and Verner, June M},
  journal={Journal of Systems and Software},
  volume={79},
  number={11},
  pages={1541--1551},
  year={2006},
  publisher={Elsevier}
}

@article{shastri2021role,
  title={The role of the project manager in agile software development projects},
  author={Shastri, Yogeshwar and Hoda, Rashina and Amor, Robert},
  journal={Journal of Systems and Software},
  volume={173},
  pages={110871},
  year={2021},
  publisher={Elsevier}
}

@inproceedings{oliveira2016software,
  title={Software project managers' perceptions of productivity factors: findings from a qualitative study},
  author={Oliveira, Edson and Conte, Tayana and Cristo, Marco and Mendes, Emilia},
  booktitle={Proceedings of the 10th ACM/IEEE international symposium on empirical software engineering and measurement},
  pages={1--6},
  year={2016}
}

@article{suwarno2025exploring,
  title={Exploring machine learning to support software managers' insights on software developer productivity},
  author={Suwarno, Suwarno and Christian, Yefta and Yoputra, Keaton and Estrada, Yuki},
  journal={Journal of Applied Data Sciences},
  volume={6},
  number={3},
  pages={1504--1524},
  year={2025}
}

@inproceedings{peters2015educating,
  title={Educating software engineering managers-revisited what software project managers need to know today},
  author={Peters, Lawrence and Moreno, Ana M},
  booktitle={2015 IEEE/ACM 37th IEEE International Conference on Software Engineering},
  volume={2},
  pages={353--359},
  year={2015},
  organization={IEEE}
}

@article{c,
  title={Software project management practices: Failure versus success},
  author={Jones, Capers},
  journal={CrossTalk: The Journal of Defense Software Engineering},
  volume={17},
  number={10},
  pages={5--9},
  year={2004}
}

@article{assalaarachchi2025generative,
  title={Generative AI for Software Project Management: Insights from a Review of Software Practitioner Literature},
  author={Assalaarachchi, Lakshana Iruni and Masood, Zainab and Hoda, Rashina and Grundy, John},
  journal={IEEE Software},
  year={2025},
  publisher={IEEE}
}

@article{ali2026artificial,
  title={Artificial intelligence and machine learning in enhancing software project management processes: A systematic literature review},
  author={Ali, Usama and Naseer, Mehwish},
  journal={Automated Software Engineering},
  volume={33},
  number={1},
  pages={31},
  year={2026},
  publisher={Springer}
}

@inproceedings{berenguer2024investigating,
  title={Investigating Entry-Level Software Project Managers’ Skills and Responsibilities: An Empirical Analysis of LinkedIn Job Ads},
  author={Berenguer, Clara and Freire, S{\'a}vio and Mendon{\c{c}}a, Manoel and Sp{\'\i}nola, Rodrigo},
  booktitle={Proceedings of the 26th International Conference on Enterprise Information Systems},
  volume={2},
  pages={214--221},
  year={2024}
}

@inproceedings{rasnayaka2024empirical,
  title={An empirical study on usage and perceptions of llms in a software engineering project},
  author={Rasnayaka, Sanka and Wang, Guanlin and Shariffdeen, Ridwan and Iyer, Ganesh Neelakanta},
  booktitle={Proceedings of the 1st International Workshop on Large Language Models for Code},
  pages={111--118},
  year={2024}
}

@article{zheng2025towards,
  title={Towards an understanding of large language models in software engineering tasks},
  author={Zheng, Zibin and Ning, Kaiwen and Zhong, Qingyuan and Chen, Jiachi and Chen, Wenqing and Guo, Lianghong and Wang, Weicheng and Wang, Yanlin},
  journal={Empirical Software Engineering},
  volume={30},
  number={2},
  pages={50},
  year={2025},
  publisher={Springer}
}

@article{hou2024large,
  title={Large language models for software engineering: A systematic literature review},
  author={Hou, Xinyi and Zhao, Yanjie and Liu, Yue and Yang, Zhou and Wang, Kailong and Li, Li and Luo, Xiapu and Lo, David and Grundy, John and Wang, Haoyu},
  journal={ACM Transactions on Software Engineering and Methodology},
  volume={33},
  number={8},
  pages={1--79},
  year={2024},
  publisher={ACM New York, NY}
}

@inproceedings{champa2024chatgpt,
  title={Chatgpt in action: Analyzing its use in software development},
  author={Champa, Arifa Islam and Rabbi, Md Fazle and Nachuma, Costain and Zibran, Minhaz F},
  booktitle={Proceedings of the 21st International Conference on Mining Software Repositories},
  pages={182--186},
  year={2024}
}

@inproceedings{meem2025software,
  title={Why do Software Practitioners Use ChatGPT for Software Development Tasks?},
  author={Meem, Fairuz Nawer and Smith, Justin and Johnson-Matthews, Brittany},
  booktitle={Proceedings of the 33rd ACM International Conference on the Foundations of Software Engineering},
  pages={1508--1514},
  year={2025}
}

@article{khojah2024beyond,
  title={Beyond code generation: An observational study of chatgpt usage in software engineering practice},
  author={Khojah, Ranim and Mohamad, Mazen and Leitner, Philipp and de Oliveira Neto, Francisco Gomes},
  journal={Proceedings of the ACM on Software Engineering},
  volume={1},
  number={FSE},
  pages={1819--1840},
  year={2024},
  publisher={ACM New York, NY, USA}
}

@inproceedings{santana2025software,
  title={Software Testing with Large Language Models: An Interview Study with Practitioners},
  author={Santana, Deolinda and Magalhaes, Cleyton and de Souza Santos, Ronnie},
  booktitle={2025 2nd IEEE/ACM International Conference on AI-powered Software (AIware)},
  pages={96--104},
  year={2025},
  organization={IEEE}
}

@inproceedings{santos2025model,
  title={Model-Assisted and Human-Guided: Perceptions and Practices of Software Professionals Using LLMs for Coding},
  author={Santos, Italo and Magalhaes, Cleyton and Santos, Ronnie De Souza},
  booktitle={2025 2nd IEEE/ACM International Conference on AI-powered Software (AIware)},
  pages={105--112},
  year={2025},
  organization={IEEE}
}

@inproceedings{coutinho2024role,
  title={The role of generative ai in software development productivity: A pilot case study},
  author={Coutinho, Mariana and Marques, Lorena and Santos, Anderson and Dahia, Marcio and Fran{\c{c}}a, Cesar and de Souza Santos, Ronnie},
  booktitle={Proceedings of the 1st ACM International Conference on AI-Powered Software},
  pages={131--138},
  year={2024}
}

@inproceedings{santos2024we,
  title={Are we testing or being tested? exploring the practical applications of large language models in software testing},
  author={Santos, Robson and Santos, Italo and Magalhaes, Cleyton and de Souza Santos, Ronnie},
  booktitle={2024 IEEE Conference on Software Testing, Verification and Validation (ICST)},
  pages={353--360},
  year={2024},
  organization={IEEE}
}

@inproceedings{krishna2024using,
  title={Using llms in software requirements specifications: An empirical evaluation},
  author={Krishna, Madhava and Gaur, Bhagesh and Verma, Arsh and Jalote, Pankaj},
  booktitle={2024 IEEE 32nd International Requirements Engineering Conference (RE)},
  pages={475--483},
  year={2024},
  organization={IEEE}
}

@incollection{cabrero2024can,
  title={How Can Generative AI Enhance Software Management? Is It Better Done than Perfect?},
  author={Cabrero-Daniel, Beatriz and Fazelidehkordi, Yasamin and Nouri, Ali},
  booktitle={Generative AI for Effective Software Development},
  pages={235--255},
  year={2024},
  publisher={Springer}
}

@inproceedings{rico2025challenges,
  title={Challenges and Opportunities for Generative AI in Software Engineering: A Managerial View},
  author={Rico, Sergio and {\"O}berg, Lena-Maria},
  booktitle={Proceedings of the 33rd ACM International Conference on the Foundations of Software Engineering},
  pages={1338--1344},
  year={2025}
}

@misc{ralph2020empirical,
      title={Empirical Standards for Software Engineering Research}, 
      author={Paul Ralph et al.}, 
      year={2020}, 
      eprint={2010.03525}, 
      archivePrefix={arXiv}, 
      primaryClass={cs.SE}, 
      url={https://arxiv.org/abs/2010.03525} }

@article{runeson2009guidelines,
  title={Guidelines for conducting and reporting case study research in software engineering},
  author={Runeson, Per and H{\"o}st, Martin},
  journal={Empirical software engineering},
  volume={14},
  number={2},
  pages={131--164},
  year={2009},
  publisher={Springer}
}

@article{terry2017thematic,
  title={Thematic analysis},
  author={Terry, Gareth and Hayfield, Nikki and Clarke, Victoria and Braun, Virginia and others},
  journal={The SAGE handbook of qualitative research in psychology},
  volume={2},
  number={17-37},
  pages={25},
  year={2017},
  publisher={SAGE Publications Ltd}
}

@incollection{george2018descriptive,
  title={Descriptive statistics},
  author={George, Darren and Mallery, Paul},
  booktitle={IBM SPSS Statistics 25 Step by Step},
  pages={126--134},
  year={2018},
  publisher={Routledge}
}

@article{seaman1999qualitative,
  title={Qualitative methods in empirical studies of software engineering},
  author={Seaman, Carolyn B.},
  journal={IEEE Transactions on software engineering},
  volume={25},
  number={4},
  pages={557--572},
  year={1999},
  publisher={IEEE}
}

@article{jones2004software,
  title={Software project management practices: Failure versus success},
  author={Jones, Capers},
  journal={CrossTalk: The Journal of Defense Software Engineering},
  volume={17},
  number={10},
  pages={5--9},
  year={2004}
}

@article{baker2012many,
  title={How many qualitative interviews is enough},
  author={Baker, Sarah Elsie and Edwards, Rosalind},
  year={2012},
  publisher={NCRM}
}

@article{elmholdt2026many,
  title={'How many interviews do I need?'An examination of interview numbers and sampling moves in qualitative research},
  author={Elmholdt, KT and Gill, MJ and Nielsen, JA},
  journal={Organizational Research Methods},
  year={2026},
  publisher={SAGE Publications}
}

@article{xia2025demystifying,
  title={Demystifying llm-based software engineering agents},
  author={Xia, Chunqiu Steven and Deng, Yinlin and Dunn, Soren and Zhang, Lingming},
  journal={Proceedings of the ACM on Software Engineering},
  volume={2},
  number={FSE},
  pages={801--824},
  year={2025},
  publisher={ACM New York, NY, USA}
}

@article{he2026llm,
  title={Llm-as-a-judge for software engineering: Literature review, vision, and the road ahead},
  author={He, Junda and Shi, Jieke and Zhuo, Terry Yue and Treude, Christoph and Sun, Jiamou and Xing, Zhenchang and Du, Xiaoning and Lo, David},
  journal={ACM Transactions on Software Engineering and Methodology},
  year={2026},
  publisher={ACM New York, NY}
}

@article{zhao2025code,
  title={From Code Completion to Autonomous Pipeline Orchestration: How LLM-Powered Developer Tools Are Reshaping Software Engineering Workflows},
  author={Zhao, Wenyu and Shang, Wenbin and Liu, Yang},
  journal={American Journal Of Big Data},
  volume={6},
  number={05},
  pages={111--139},
  year={2025}
}

@article{zhong2025large,
  title={Large Language Models in Software Engineering: Automation, Collaboration, and Challenges},
  author={Zhong, Cheng},
  journal={Advances in Engineering Technology Research},
  volume={15},
  number={1},
  pages={1795--1795},
  year={2025}
}

@article{guest2006many,
  title={How many interviews are enough? An experiment with data saturation and variability},
  author={Guest, Greg and Bunce, Arwen and Johnson, Laura},
  journal={Field methods},
  volume={18},
  number={1},
  pages={59--82},
  year={2006},
  publisher={Sage Publications Sage CA: Thousand Oaks, CA}
}
